\documentclass[a4paper]{jpconf}

\usepackage{graphicx}
\usepackage{amsmath}
\usepackage{amssymb,amsmath}
\usepackage{mathrsfs}
\usepackage{journal_names_saip}

\newcommand{\kph}{km\,s$^{-1}$}
\newcommand{\dg}{^{\circ}}

\def\la{\mathrel{\hbox{\rlap{\hbox{\lower4pt\hbox{$\sim$}}}\hbox{$<$}}}}
\def\ga{\mathrel{\hbox{\rlap{\hbox{\lower4pt\hbox{$\sim$}}}\hbox{$>$}}}}

\def\arcmin{\hbox{$^\prime$}}

\def\fm{\hbox{$.\!\!^{\rm m}$}}

\def\fdg{\hbox{$.\!\!^\circ$}}

\newcommand{\kms}{{\,km\,s$^{-1}$}}

\newcommand{\HI}{\mbox{\normalsize H\thinspace\footnotesize I}}

\def\apjl   {{ApJL }}
\def\apjs   {{ApJS }}

\begin{document}
\title{Extragalactic large-scale structures in the northern Zone of Avoidance}


\author{M Ramatsoku$^{1}$, R C Kraan-Korteweg$^{1}$, A C Schr\"{o}der$^{2}$ and W van
Driel$^{3}$}

\address{$^{1}$Astrophysics, Cosmology and Gravity Centre (ACGC), Department of Astronomy,
University of Cape Town, Private Bag X3, Rondebosch 7701, South Africa.}
\address{$^{2}$South African Astronomical Observatory (SAAO), PO Box 9, 7935 Observatory, Cape Town,
South Africa.}
\address{$^{3}$GEPI, Observatoire de Paris, CNRS, Université Paris Diderot, 5 place Jules Janssen,
92190 Meudon, France.}

\ead{mpati.ramatsoku@ast.uct.ac.za, kraan@ast.uct.ac.za, anja@hartrao.ac.za, wim.vandriel@obspm.fr}

\begin{abstract}
  We used the Nan\c{c}ay Radio Telescope (NRT) to measure the 21\,cm line
  emission of near-infrared bright galaxies in the northern Zone of
  Avoidance (ZoA) without previous redshift determinations. We
  selected galaxies with extinction-corrected magnitudes $K _{s}^{o}
  \le 11\fm25$ from the 2MASS Extended Source Catalog. These data will
  complement the existing 2MASS Redshift Survey (2MRS; first data
  release) as well as the ongoing 2MASS Tully-Fisher survey, both of
  which exclude the inner ZoA ($|b|< 5^{\circ}$), where the
  identification of galaxy candidates is the hardest. Of the
  $\sim$1000 identified 2MASX galaxy candidates we have so far
  detected 252 to our 3.0 mJy rms sensitivity limit and the velocity
  limit of 10500~km~s$^{-1}$. The resulting redshift distribution
  reveals various new structures that were hitherto uncharted. They
  seem to form part of the larger Perseus-Pisces Supercluster (PPS). The
  most conspicuous is a ridge at about $\ell\approx~160^{\circ}$,$v
  \approx 6500$\,\kms.  Within this wall-like
  structure, two strong radio galaxies (3C\,129 and 3C\,129.1) are
  embedded which lie at the same distance as the ridge. They seem to
  form part of an X-ray cluster. Another prominent filament has been
  identified crossing the ZoA at $\ell \approx 90^\circ$, hence
  suggesting the second Perseus-Pisces arm is more extended than
  previously thought.
\end{abstract}

\section{Introduction}
Dust extinction and high stellar densities in the Galactic Plane block
a large fraction of the sky resulting in the so-called Zone of
Avoidance (ZoA) \cite{kraan05}. Compared to the optical, the
near-infrared (NIR) is much less affected by the dust obscuration. A
whole-sky near-infrared ($JHK$) imaging survey exists in the form of
the 2-Micron All Sky Survey (2MASS)\cite{Strustskie2006}. The
resulting extended source catalogue, 2MASX, with its 1.6 million
sources complete to $K_{s} \le 13\fm5$ \cite{jarrett2003}, provides
the most uniform and deep NIR sample of the whole sky. Although 2MASX
suffers little from dust extinction, there remains an ``NIR ZoA''
caused by stellar crowding around the Galactic bulge ($\ell \la \pm$
90$\dg$) \cite{kraan05}.

To analyse the large-scale galaxy distribution over the entire sky,
the optical 2MASS Redshift Survey (2MRS) was started about a decade
ago. The first data release is complete to $K _{s}^{o} = 11\fm25$
\cite{huchra2005}, the second to $K _{s}^{o} = 11\fm75$
\cite{huchra2012}. Both versions do exclude, however, the inner ZoA
($|b| \le 5\dg$) because of the inherent difficulties in getting good
signal-to-noise (SNR) optical spectra for these heavily obscured
galaxies. While the 2MRS is currently the deepest "whole-sky" redshift
survey for mapping large-scale structures, and studying the dynamics
in the nearby Universe and the CMB dipole \cite{kraan00}, the lack of
redshift data in the ZoA remains an obstacle. This also holds for the
2MASS Tully Fisher survey (2MTF) which uses a subsample of
sufficiently inclined 2MASX spiral galaxies to study cosmic flow
fields \cite{masters2008}.

To improve on this we have started a project to systematically observe
in \HI\ all likely 2MASX galaxies in the ``2MRS ZoA'' without previous
redshift information. The line emission from neutral hydrogen (\HI) at
the radio wavelength of 21 cm can travel unhindered through the
thickest dust layers of the Milky Way. Targeting ZoA galaxies with a
radio telescope will therefore allow us to obtain redshifts for
gas-rich 2MASS galaxies. We used the 100m-class Nan\c{c}ay Radio
Telescope (NRT) for pointed observations of all ZoA galaxies with Dec
$> -39\dg$. We were particularly interested in filling in the northern
ZoA because -- contrary to the southern ZoA -- most of the northern
ZoA has not been surveyed in any systematic way before. For the
southern hemisphere a blind \HI\ survey has been performed with the
Parkes Multi-Beam receiver (HIZOA) which covers $|b| < 5\dg$ for the
longitude range $196\dg \le \ell \le 52\dg$ out to velocities of
12700\,km~s$^{-1}$ (rms~$\approx~6$ mJy)\cite{henning10},
\cite{Donley2005}. For the remaining ZoA ($196\dg \ga \ell \ga
52\dg$) hardly any data are available (see e.g. the top panel of
Fig.\ref{LSS}).

\section{Filling in the redshift gap in the ZoA}

\subsection{Sample selection}
To start filling in the northern redshift gap we first extracted all
extended sources from 2MASX with $|b| \le 10\dg$ and
extinction-corrected magnitudes $K_{s}^{o} \le 11\fm25$, i.e., the
completeness limit of the first 2MRS catalogue and the 2MTF. Of the
4743 extended sources accessible to the NRT (Dec $> -39\dg$), we classified 2546 sources as clear galaxies (plus
42 as possible galaxies) by visually inspecting the Digitized Sky
Survey (DSS)\footnote[1]{http://stdatu.stsci.edu/dss/} images in the
$B_J$, $R$ and $I$ bands, the 2MASS $J$ and $K$-bands, as well as
the 2MASX colour images. From that sample we excluded galaxies that
already had redshift measurements by cross-correlating our catalogue
with NED and unpublished data sets like 2MRS (Macri, priv comm), HIZOA
(Kraan-Korteweg et al, priv comm), etc.

The final NRT target sample consists of $\sim$1000 near-infrared
bright galaxy candidates in the ZoA ($-20^{\circ} \la \ell \la
270^{\circ}$; $|b| \le 10^\circ$); the great majority ($\ga$ 83\%)
of them are located in the $|b| < 5^\circ$ strip. We have already used
1200 hours of observing time with the NRT for pointed observations of
these objects since mid-2009.

\subsection{Observations and data reduction}
The galaxy candidates were observed with the NRT in position-switching mode. Pairs of equal-duration
ON/OFF-source integrations were made, with the OFF-source position
20${\arcmin}$ east of the target. Candidates were typically observed
for 40-minute long periods till an rms noise level of 3.0 mJy was
reached. Because none of the target galaxies have prior redshift
information, the auto-correlator was set to cover a radial velocity
range of $-$500 to 10500~km~s$^{-1}$. The original resolution is 2.6
\kph, which is later smoothed to 18 \kph\ for further analysis. Observations were made simultaneously in two linear
polarizations, to gain sensitivity. Data reduction and Radio Frequency Interference (RFI) recognition
and mitigation were performed using the NAPS and SIR data reduction
packages, developed by the NRT staff.

\subsection{Results}
From July 2009 to March 2012, we observed 926 out of the total of 1000
target galaxies to an rms level of 3.0~mJy.
The resulting spectra were first inspected by eye by one of us for the
signature of redshifted Galactic \HI-emission. The reliability of all
potential \HI\ detections was then assessed independently by three
other team members, followed by an adjudication by another team
member. Clear detections, and non-detections which had reached our
target rms noise level of 3.0 mJy were filed as such, whereas marginal
or possible detections were tagged for re-observation, whose results
were continuously updated.

This has led to 252 solid detections so far out of the 926 observed
targets, i.e., a detection rate of 27\%. This is a respectable
detection rate given that no pre-selection was made according to
morphological type (which is not straightforward in the NIR, nor in
the ZoA). The mean rms in the final spectra was found to be 2.9
mJy. The 252 detected \HI\ profiles have a peak
signal-to-noise ratio $\ga 5.0$.

A reliable detection was found to have a typical linewidth- and flux-dependent signal-to-noise ratio as defined in the  Arecibo Legacy Fast ALFA survey of SNR$_{ALFALFA}> 6.0$ \cite{Saintonge2007}. Detections with
$5.0 \la {\rm SNR}_{ALFALFA} \la 6.0$ typically were adjudicated as marginal.

\section{Resulting large-scale structures}
While we find galaxies over the entire observed redshift range, the
majority lie within $2000 - 8000$\kms. The velocity histogram (not
shown here) shows a clear peak at 6000\kms, probably due to the
prevalence of galaxies connected to the Perseus-Pisces Supercluster
(PPS) described below.  There is a noticeable drop-off in detections around $\sim 8000$\kms\ which is due to a combination of
recurring RFI at $v > 8500$\kms\ and limited telescope sensitivity.

The new \HI\ detections are distributed almost symmetrically about the
Galactic equator, irrespective of Galactic latitude $b$ (see the
middle panel of Fig.~1), confirming that the
detection rate is independent of extinction and star density.

\begin{figure}[htp!]
\begin{center}
\includegraphics[width=14cm,height=12cm]{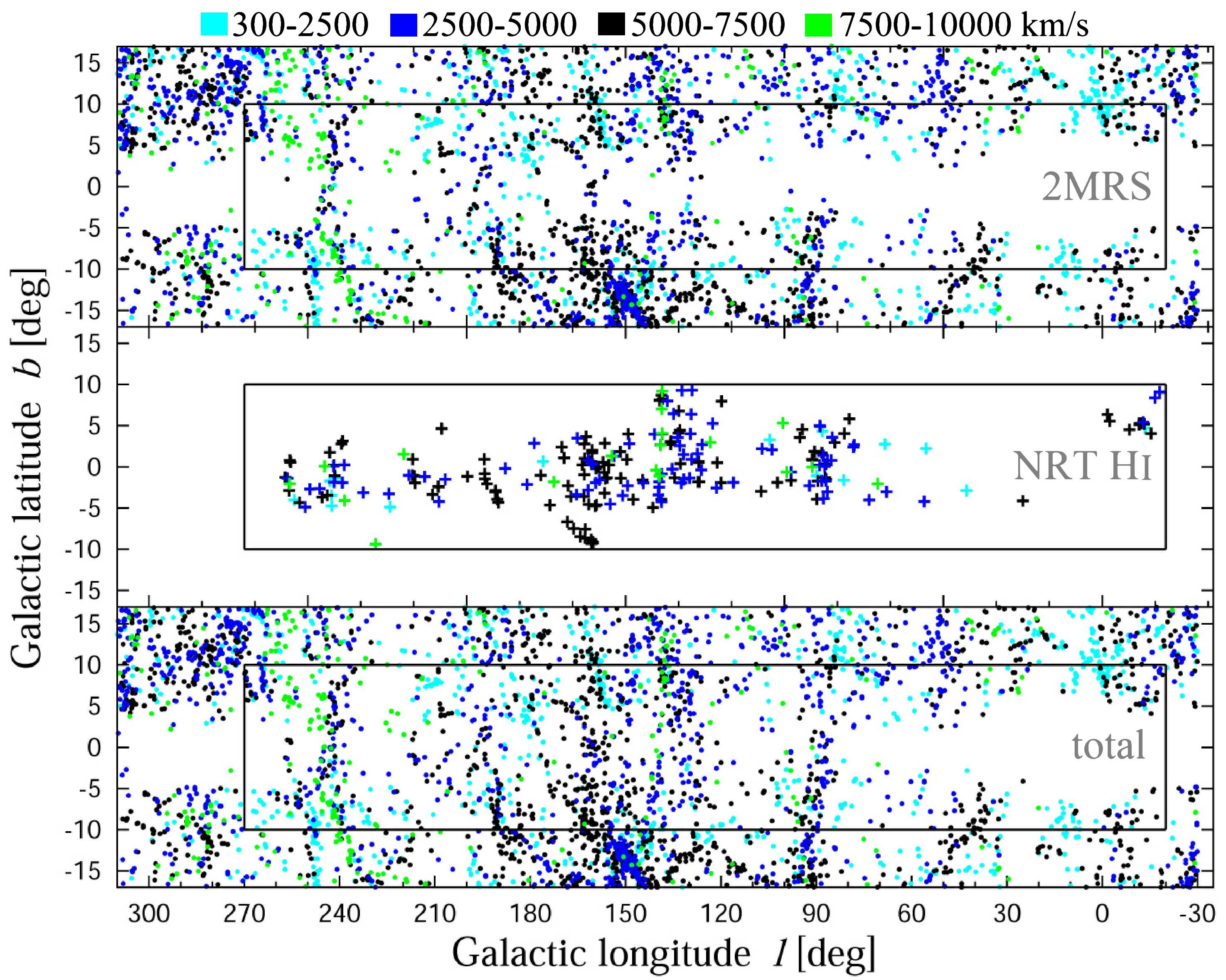}
\end{center}
\caption[]{The spatial distribution of galaxies in Galactic
  coordinates in the ZoA, showing large-scale structure filaments and voids. The survey area explored with the NRT is
  marked by the black rectangle. Galaxies are colour-coded by their
  velocity range, as shown at the top: the $300-2500$\kms\ velocity
  range is shown in cyan, $2500-5000$\kms\ in blue, $5000-7500$\kms\
  in black and the $7500-10\,000$\kms\ range in green. The top panel
  shows the 2MRS galaxies ($K_{s}^{o} \le 11\fm25$) with known
  redshifts from Zcat (Huchra, priv comm) for low-latitude 2MASX
  galaxies. The middle panel displays the \HI-detections obtained by
  us with the NRT. The bottom panel shows the combined data and
  emphasizes the links with previously known structures.}\label{LSS}
\end{figure}

To investigate the large-scale structures revealed by these new
detections we plot in Fig.~\ref{LSS} their spatial distribution in
Galactic coordinates centred on the northern Milky Way. It includes
2MRS data up to latitudes of $|b| \ge 15\dg$ to test for continuity of
the newly identified features with previously known structures
(filaments, walls, voids) at higher Galactic latitudes. The outlined
rectangular region demarcates the NRT survey area.

The top panel shows the 2MRS galaxies with $K _{s}^{o} \le 11\fm25$,
as well as data in the $|b| \la 5\dg$ strip from the Zcat
($K_{s}^{o} \le 11\fm25$) compilation by Huchra (priv comm). The
middle panel displays the distribution of the 252 NIR-bright 2MASX
galaxies detected with the NRT.  In the bottom panel the new
detections have been merged with previously known data of objects
to the $K _{s}^{o} \le 11\fm25$ completeness limit. In all three
panels galaxies are colour-coded by velocity: the $300-2500$\kms\
velocity range is shown in cyan, $2500-5000$\kms\ in blue,
$5000-7500$\kms\ in black and the $7500-10000$\kms\ range in
green. The black and blue colours coincide with the approximate
velocity range of the PPS.

When comparing the bottom and top panels (i.e., the status before and
after the NRT observations) the power of revealing previously unknown
large-scale structures in the ZoA through \HI\ observations of
intrinsically bright (extinction-corrected) NIR galaxies is quite
obvious. Several prominent filaments and walls are now seen to cross
the ZoA which were previously not -- or at most marginally --
visible. The most obvious filamentary structures cross the Galactic
Plane at $\ell \approx 90^{\circ}$ (Cygnus), $\ell \approx
135^{\circ}$ (Cassiopeia), $\ell \approx 160^{\circ}$ (Perseus), $\ell
\approx 180^{\circ}$, as well as the Puppis filament at $\ell
\approx 240^{\circ}$ \cite{Kraan1992}. An underdense region of
galaxies is apparent at $\ell \approx 120^{\circ}$ and
$-2\dg \la b \la 5\dg$, stretching from $cz \approx 4000$ to
7000\kms. The 2MASX finds no galaxies in regions of high stellar densities, such as around the Galactic Bulge, therefore the ZoA  for $\ell \la 30\dg$ remains unsampled. In the next sections we will discuss only two of the most
striking newly revealed features in more detail, namely the ones at
$\ell \approx 90^{\circ}$ and $\ell \approx 160^{\circ}$.

\newpage
\subsection{The Perseus extension of the second Perseus-Pisces arm}
The first region of interest is seen in the constellation of Cygnus
around $\ell$ $\approx$ 90$^{\circ}$. It shows a very prominent
filament that can be traced from below the Plane at $\ell,b \approx
90\dg,-10\dg$ extending up to the other side of the ZoA at a slight
angle to $\ell,b \approx 100\dg,-10\dg$. This filament seems to form
part of the second (eastern) PPS arm that emanates
southwards from the Perseus A\,426 cluster ($\ell$,\textit{b})
$\approx$ (150$^\circ$,-13$^\circ$), then bends backwards towards the
Galactic plane (at about $\ell,b\approx 110\dg,-30\dg$; not shown
here) and re-enters the plot at about $\ell,b \approx
80\dg,-15\dg$. Most previous studies of the PPS assumed it to kind
of stop and dissolve around $\ell,b \approx 90\dg,-10\dg$ as no
signature was found of it in earlier optical galaxy searches
\cite{seerbeger94}, nor any indication of a continuation on the other
side of the obscuring ZoA band.  Our data clearly confirms such a
continuation. It implies the eastern Perseus-Pisces chain to be
considerably larger than evidenced in any previous survey of the PPS
complex.

\subsection{A potentially massive cluster}\label{massiveCluster}
The second prominent feature is a concentration of \HI-detections at
$\ell \approx 160^\circ$ right in the middle of the ZoA ($b =
0\fdg5$). It lies within a nearly vertical (in Fig.~1) wall-like
structure (at $v \approx 6000 \pm 1000$\kms\ in velocity space) and
can be traced across the full width of the ZoA. It is interesting to
note that within this wall, at the core of the galaxy concentration,
we find two very strong radio galaxies. Their position and redshifts
(6236 and 6655~km~s$^{-1}$ respectively; \cite{Spinrad1975}) confirm
that they reside inside the galaxy concentration. These are the
head-tail radio source (3C\,129) and the double-lobed giant elliptical
radio galaxy (3C\,129.1). The presence of such radio sources with
bent lobe morphology usually is indicative of a rich cluster
environment.

Focardi et al (1984) \cite{focardi84} were the first to put forward
the idea of an extension of the PPS Complex across the ZoA
towards the northern Galactic hemisphere that would link the Perseus
cluster (A\,426; $\ell,b,v \simeq 150^\circ, -13^\circ, 5000$\kms) to
Abell~569 ($\ell,b,v \simeq 168^{\circ}, 23^{\circ}, 5800$\kms). It
would traverse the Galactic Plane at the location of the two bright
radio galaxies, which also coincides with the location where
Weinberger (1980) (\cite{weinberger1980}) found an excess of galaxies
in his early optical search (at $\ell \approx 160^{\circ}$). This
connection has been much debated over the years (e.g.,
\cite{chamaraux1990}, \cite{lu95}, \cite{pantoja1997}), but no
conclusive results were found due to the lack of (redshift) data in
this dust-enshrouded region.

The suspicion that 3C\,129 and 3C\,129.1 form part of a massive
cluster was later substantiated through the identification of the
X-ray cluster CIZA~J0450.0+4501 (\cite{Ebling2000}). The radio sources
lie within the X-ray emission of the CIZA cluster and are at the same
distance. With an X-ray luminosity of $L_{\rm X}= 1.89 \times
10^{44}$~h$_{50}^{-2}$erg~s$^{-1}$ this cluster is not among the
brightest X-ray sources in ROSAT. For comparison, its flux is about
20\% that of the Norma cluster A 3627, the central cluster of the
Great Attractor \cite{kraan96} -- which, as an aside, also hosts 2
radio sources, the central one a wide-angle tail source and the other
also a head-tail source. With regards to the X-ray flux it should be
noted, however, that the intervening high gas column density ($N_{H}
\ga 10^{21}$~cm$^{-2}$) in the Galaxy may well have reduced the flux
of the low energy X-ray photons in the ROSAT $0.1 - 2.4$ keV band,
resulting in an underestimate of its luminosity. The cluster might
therefore be more massive than its X-ray luminosity suggests. Despite
its possible connection with the wider PPS complex, it has not
received much attention since.

\section{Conclusions and future perspectives}
The 252 \HI\ detections in this previously unexplored northern region
of the ZoA have revealed new and interesting structures that are
clearly associated with the Perseus Pisces Supercluster. These new
structures at $\ell$ $\approx$ 90$^{\circ}$ and $\ell$ $\approx$ 160$^{\circ}$ seem to imply that the PPS is more extended than
previously thought and potentially much more massive, with the newly
identified cluster in the $\ell$ $\approx$ 160$^{\circ}$
filament described in Sect.~\ref{massiveCluster}. This may well have implications for our understanding of the
dynamics and flow-fields observed in this region, such as the tug of war between the PPS and the Great Attactor \cite{Scharf1992, vdweygaert2000, erdogdu2006} .

To learn more about the $\ell$ $\approx$ 160$^{\circ}$ cluster's role in, and its mass contribution
to the PPS, and its relation to the observed local flow fields, we recently put in a proposal -- and have been allocated time -
to conduct a deep \HI\ imaging survey over a $2\fdg4 \times 2\fdg4$
area (mosaicked) around this cluster with the Westerbork Synthesis
Radio Telescope (WSRT).

\section*{Acknowledgements}
This work is based upon research supported by the National Research
Foundation and Department of Science and Technology. MR is grateful
for the bursary provided by the South African SKA Project Office. This
publication makes use of data products from the Two Micron All Sky
Survey, which is a joint project of the University of Massachusetts
and the Infrared Processing and Analysis Center, funded by the
National Aeronautics and Space Administration and the National Science
Foundation.

\section*{References}
\bibliographystyle{iopstyle}
\bibliography{references_saip}
\end{document}